\documentclass[aps,prl,twocolumn,groupedaddress,
showpacs,amsmath,amssymb]{revtex4}

\usepackage{graphicx}
\usepackage{dcolumn}
\usepackage{bm}

\begin{document}

\title{Ising-like Spin Anisotropy and Competing Antiferromagnetic -
Ferromagnetic Orders in GdBaCo$_{2}$O$_{5.5}$ Single Crystals}

\author{A. A. Taskin}
\altaffiliation[On leave from ]{Institute of Semiconductor
Physics, Siberian Branch of RAS, 630090 Novosibirsk, Russia.}
\author{A. N. Lavrov}
\author{Yoichi Ando}

\affiliation{Central Research Institute of Electric Power Industry,
Komae, Tokyo 201-8511, Japan}

%\date{\today}

\begin{abstract}

In $R$BaCo$_{2}$O$_{5+x}$ compounds (R is rare earth), a
ferromagnetic-antiferromagnetic competition is accompanied by a giant
magnetoresistance. We study the magnetization of detwinned
GdBaCo$_{2}$O$_{5.5}$ single crystals, and find a remarkable uniaxial
anisotropy of Co$^{3+}$ spins which is tightly linked with the chain
oxygen ordering in GdO$_{0.5}$ planes. Reflecting the underlying oxygen
order, CoO$_2$ planes also develop a spin-state order consisting of
Co$^{3+}$ ions in alternating rows of $S=1$ and $S=0$ states. The magnetic
structure appears to be composed of weakly coupled ferromagnetic ladders
with Ising-like moments, which gives a simple picture for magnetotransport
phenomena.

\end{abstract}

\pacs{75.47.Pq, 75.47.De, 75.30.Cr, 75.30.Gw}

\maketitle

Transition-metal oxides exhibit a complex interplay between charge, spin,
orbital and lattice degrees of freedom, which is at the heart of many
fascinating phenomena such as colossal magnetoresistance (MR) in
manganites. Phenomenologically, the colossal MR originates from a
magnetic-field induced transition between delicately balanced and
competing phases that dramatically differ in resistivity \cite{CMR_rev}.
However, the microscopic origin of these competing phases still remains
far from being clear, owing to the complexity of the manganites; for
example, magnetic fields (which induce a ferromagnetic (FM) spin alignment
in manganites) have been found to melt the charge order \cite{melting},
change the orbital order with accompanying Jahn-Teller distortions
\cite{Nojiri}, and modify the scale and topology of domains in
microscopically heterogeneous phases \cite{CMR_rev,inhomo}.

In order to clarify the MR mechanisms operating in transition-metal
oxides, one may look for compounds that also exhibit competing phases and
giant MR, but possess fewer degrees of freedom. Recently, an intriguing
antiferromagnetic (AF) - ferromagnetic competition accompanied with a
giant MR has been found in cobalt-oxide compounds RBaCo$_{2}$O$_{5+x}$
(where R is rare earth) \cite {Martin,Troy,Respaud,Akahoshi}. While
cobaltites with a variable spin state of Co-ions \cite {Martin, Troy,
Respaud, Akahoshi, Vogt, Moritomo, Korotin} are in general no simpler than
manganites, RBaCo$_{2}$O$_{5+x}$ with $x\approx 0.5$ possesses a plenty of
attractive features that may turn it into a model system. For $x\approx
0.5$, oxygen ions in RO$_x$ layers order into alternating filled and empty
rows running along the $a$ axis \cite {Martin, Moritomo, Akahoshi}, thus
allowing one to deal with almost perfect crystal structure. No
considerable structural distortions are detected at the AF-FM transition
as well \cite{Moritomo,Fauth}. Moreover, at $x\approx 0.5$ the valence
state of Co-ions approaches 3+, so that charge ordering and electronic
phase separation typical for mixed-valence systems \cite{CMR_rev} become
irrelevant for this composition.

In this Letter, we present a first study of the static magnetization in
detwinned GdBaCo$_{2}$O$_{5.5}$ single crystals. We find a strong
Ising-like spin anisotropy which is clearly manifested in the FM phase,
where Co spins are aligned along the oxygen-chain direction with a net
magnetic moment of $\sim 1$ $\mu_{\text{B}}/$Co. This value indicates a
1:1 ratio of Co$^{3+}$ ions in the $S=0$ and $S=1$ spin states; based on
this result and a recent structural study \cite{Frontera}, we discuss that
the oxygen chain ordering causes the $S=1$ Co$^{3+}$ ions to be arranged
into ferromagnetic 2-leg ladders that are separated by non-magnetic
layers. The AF-FM competition and giant MR in GdBaCo$_{2}$O$_{5.5}$ appear
to reflect a relative magnetic order among weakly coupled FM ladders,
which can be rather easily altered by temperature or magnetic fields.

Using the floating-zone technique, we have succeeded in growing
high-quality GdBaCo$_{2}$O$_{5+x}$ single crystals suitable for
magnetization and transport measurements. The crystals were annealed in a
flow of oxygen at 470$^{\circ}$C, the temperature being carefully tuned
\cite{unpubl} to provide the oxygen stoichiometry $x=0.50$. Parallelepiped
samples with all faces adjusted to the crystallographic planes with a
1$^{\circ}$-accuracy were prepared by cutting and polishing under Laue
x-ray diffraction control. In GdBaCo$_{2}$O$_{5.5}$, the oxygen ordering
in GdO$_{0.5}$ planes induces a tetragonal-to-orthorhombic transition,
which is accompanied by heavy twinning of crystals that mixes the $a$ and
$b$ orthorhombic axes. To detwin crystals, we slowly cooled them under a
uniaxial pressure of $\sim 0.15$ GPa from $260^{\circ}$C, using an optical
microscope to control the twin removing \cite{twins}; note that these
crystals are fragile and only few survive the detwinning procedure.
According to X-ray measurements, the remaining fraction of misoriented
domains (which characterizes the quality of detwinning) was 4-5\%.
Magnetization measurements were carried out using a SQUID magnetometer at
fields up to 7 T applied along $a$, $b$ or $c$ axis. Throughout this
paper, the magnetization coming from Co ions is determined by subtracting
the contribution of Gd-ions, assuming their ideal paramagnetic (PM)
behavior with total spin $S=7/2$ [see inset in Fig. 1(a)]; the latter is a
good approximation since no ordering of Gd$^{3+}$ moments is detected down
to 1.7 K. The magnetoresistance was measured by a four-probe method on
twinned crystals.

\begin{figure}
%\leftskip2pt
\includegraphics*[width=20.5pc]{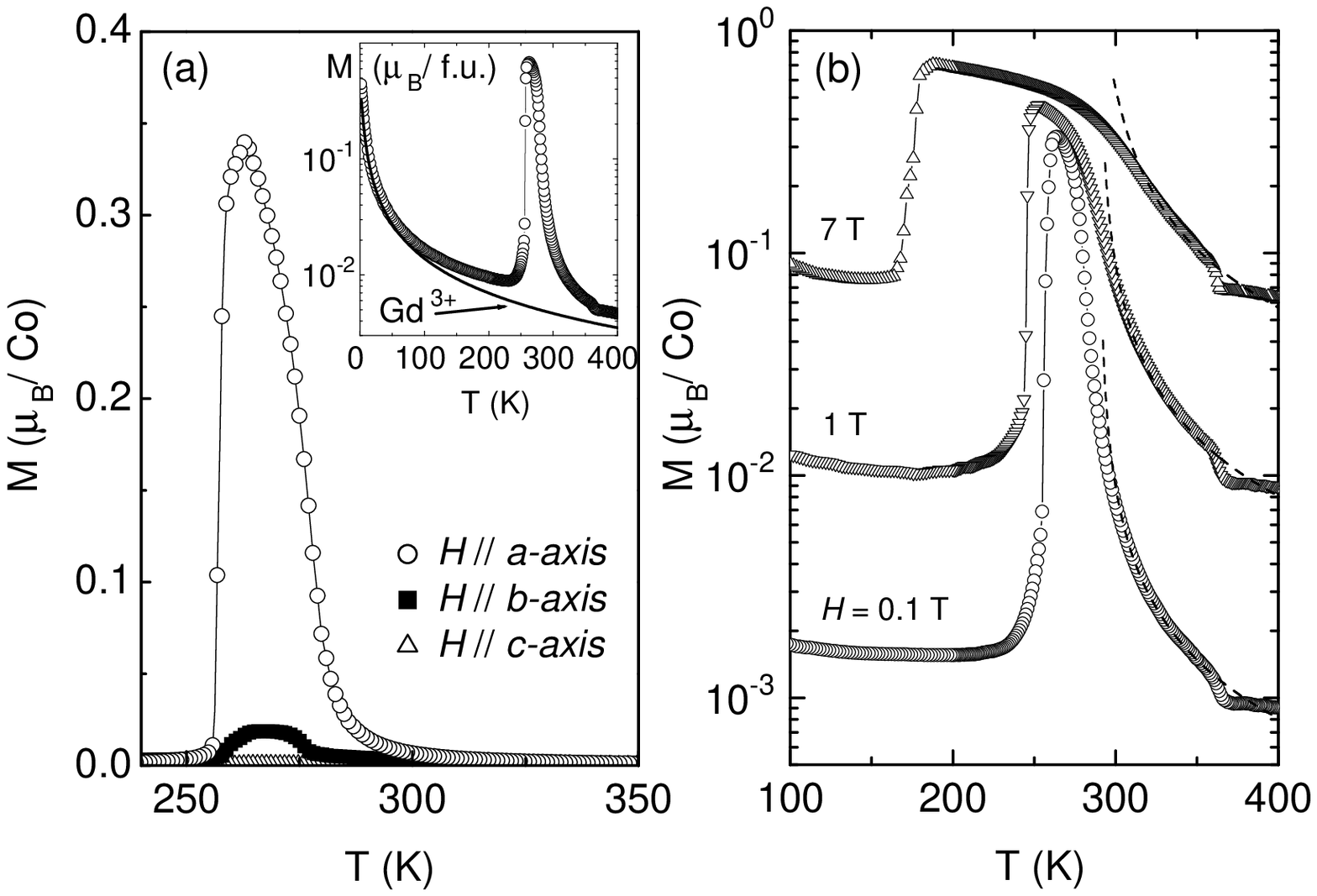}
\caption{(a) Magnetization of an untwinned GdBaCo$_{2}$O$_{5.5}$ crystal
measured in $H=0.1$ T applied along one of the crystal axis (contribution
of Gd$^{3+}$ ions is subtracted). Inset: raw $M(T)$ data for $H\parallel
a$, where the solid line shows the Curie-Weiss contribution of Gd$^{3+}$
ions ($\mu_{\text{eff}}=7.94$ $\mu_{\text{B}}$; $\theta = 0$ K). (b)
$M(T)$ in different magnetic fields $H\parallel a$. Kinks at $T\approx
360$ K correspond to the metal-insulator transition. Dashed lines are
Curie-Weiss curves with $\mu_{\text{eff}} = 2$ $\mu_{\text{B}}/$Co and
$\theta = 290$ K.}
\label{fig1}
\end{figure}

Figure 1(a) shows the magnetization of untwinned GdBaCo$_2$O$_{5.5}$
single crystals measured along $a$, $b$, and $c$ axes in the field-cooling
(FC) process. Below 300 K, a net ferromagnetic component appears in the Co
sublattice and suddenly vanishes at $\sim 260$ K, indicating successive
PM-FM-AF transitions \cite {Martin,Troy,Moritomo,Respaud,Akahoshi}. A
remarkable feature of the FM state is not only the narrow temperature
window where it shows up, but also a very strong anisotropy: We have found
that the net FM moment appears only along the $a$ axis, and even the 7-T
field is not enough to turn over the magnetic moment along the $b$ or $c$
axis. Note that a small magnetization along the $b$ axis in Fig. 1(a)
comes mostly from residual misoriented domains. This behavior suggests the
spin system in GdBaCo$_{2}$O$_{5.5}$ to be {\it Ising-like}, which
drastically simplifies the understanding of magnetic ordering.

The balance of FM and AF ordering in GdBaCo$_{2}$O$_{5+x}$ turns out to be
quite delicate, and can be easily affected by temperature, magnetic
fields, or even subtle variation in stoichiometry. Magnetic fields applied
along the spin-easy $a$ axis stabilize the FM state and shift the FM-AF
transition to lower temperatures [Fig. 1(b)]. Nevertheless, it would not
be correct to consider the FM and AF orders as simply competing ones. If
this switching were originating from equally strong and competing AF and
FM exchange interactions, spin fluctuations would inevitably be enhanced
in the vicinity of the FM-AF phase boundary, smearing the transition.
Isothermal magnetization curves in Fig. 2(a) demonstrate, however, that
although the FM-AF balance is subtle, and fairly weak fields ($\sim 1.5$ T
at 240 K) are capable of recovering the FM order, the AF-FM transition
remains always sharp, showing that thermal fluctuations are irrelevant
here. This behavior clearly indicates that the observed AF-FM switch
occurs {\it within} the ordered spin state and is governed by
reorientation of spins in one of weakly-coupled sublattices.

\begin{figure}[t]
\includegraphics*[width=20.5pc]{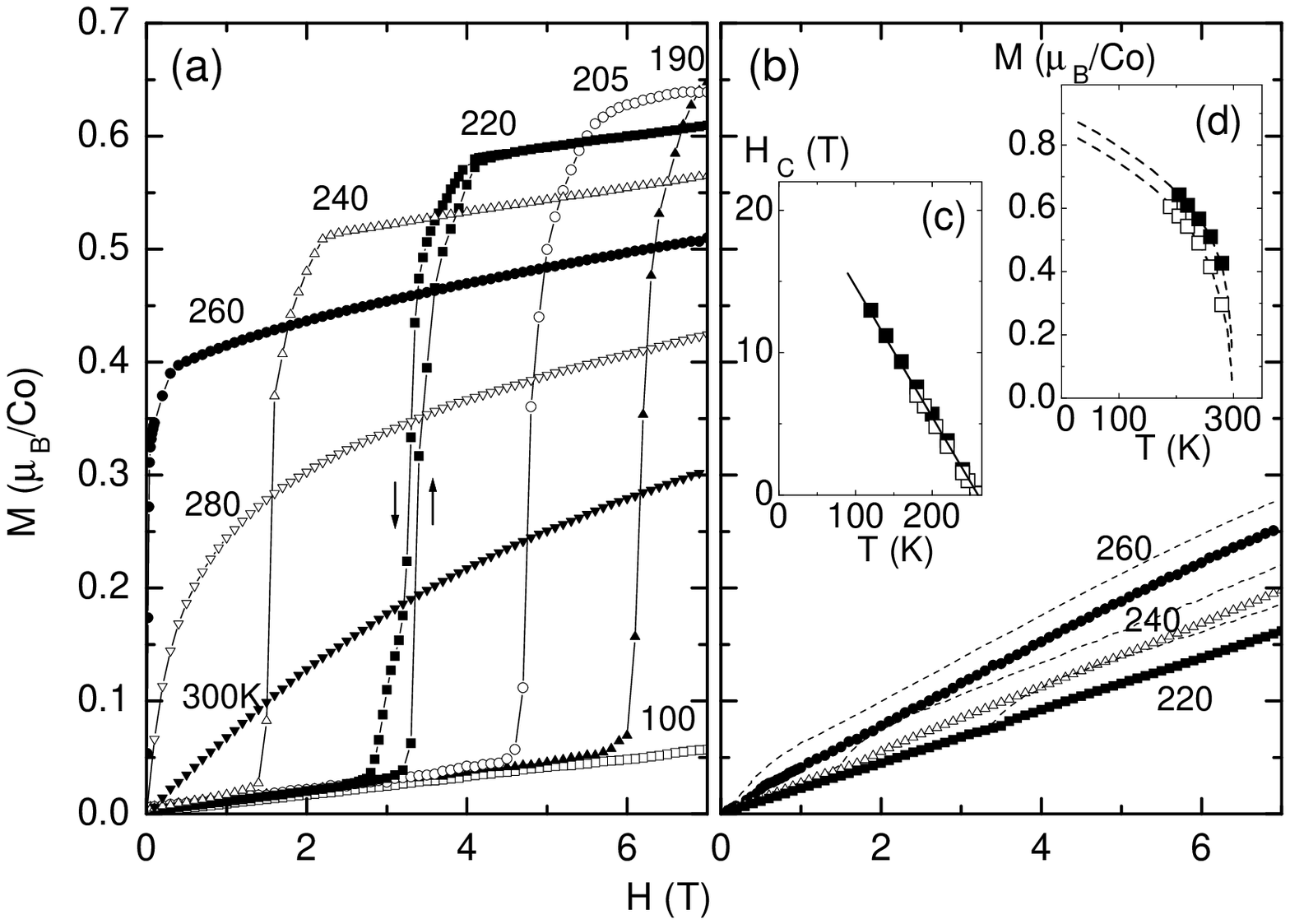}
\caption{Isothermal magnetization of GdBaCo$_{2}$O$_{5.5}$ for
(a) $H\parallel a$, (b) $H\parallel b$. In panel (b), dash lines indicate
the experimental data, and symbols show $M(H)$ after subtracting the
contribution from misoriented domains (4\% of total amount). Inset (c):
critical magnetic fields $H_c$ for the AF-FM transition obtained from
magnetization (open squares) and magnetoresistance (solid squares). Inset
(d): net magnetic moment along the $a$ axis measured at $H=7$ T (solid
squares), and the extrapolated FM moment at $H=0$ (open squares). Dashed
lines are a guide to the eye.}
\label{fig2}
\end{figure}

The fact that the AF-FM transition remains sharp, even when it is induced
by a weak field $\mu_{\text{B}}H\ll kT$ [Fig. 2(a)], unambiguously points
to a strong hierarchy of spin interactions in GdBaCo$_{2}$O$_{5.5}$. A
strong ferromagnetic interaction aligns spins within each sublattice,
while a much weaker antiferromagnetic (at $T<260$ K) interaction provides
a subtle coupling between the sublattices, which can be broken by an
applied field. Figure 2(c) shows that a magnetic field $H\parallel a$
required to overcome the AF coupling grows roughly linearly upon cooling,
from zero at $T\approx 260$ K up to $\sim 20$ T at $T=0$ (Fig. 2(c)).
Whatever the temperature, however, the 7-T field $H\parallel b$ or
$H\parallel c$, cannot compete with the spin anisotropy and causes just a
partial tilting of spins (in each sublattice) from their easy axis, thus
giving linear $M(H)$ curves in both FM and AF regions (Fig. 2(b)).

Qualitatively, the overall magnetic behavior is quite reminiscent of that
in canted antiferromagnets, such as La$_2$CuO$_4$ \cite{LCO}, where slight
spin canting brings about a weak ferromagnetism. However, the magnitude of
the field-induced FM moment, reaching $\sim 0.6$ $\mu_{\text{B}} /$Co at
$T=205$ K [Fig. 2(a)], is apparently inconsistent with such weak spin
canting, and points to a different origin of the ferromagnetism. A rough
extrapolation to $T=0$ suggests a saturated magnetic moment of $\sim 1$
$\mu_{\text{B}}/$Co [Fig. 2(d)], which corresponds to a 1:1 mixture of
low-spin (LS:~~$t^6_{2g}$, $e^0_g$; $S=0$) and intermediate-spin
(IS:~~$t^5_{2g}$, $e^1_g$; $S=1$) states of Co$^{3+}$ ions, if a simple FM
spin order is realized. A similar conclusion, that Co$^{3+}$ ions exhibit
a 1:1 ratio of LS and IS states below the metal-insulator transition at
$T\approx 360$ K \cite{Respaud, Frontera, Martin}, can be reached based on
the Curie-Weiss fitting of the PM susceptibility in the temperature range
300-360 K [Fig. 1(b)]. It is worth noting that {\it polycrystalline}
RBaCo$_{2}$O$_{5.5}$ samples demonstrate smaller FM moments
\cite{Troy,Respaud,Akahoshi}, seemingly inconsistent with the expected Co
spin states, which has been one of the mysteries of the FM state. The
discovered Ising anisotropy, that prevents moments from being seen along
the $b$ and $c$ axes, readily resolves this discrepancy.

To understand why Co$^{3+}$ ions in GdBaCo$_{2}$O$_{5.5}$ even at zero
temperature exhibit two different spin states, one should consider the
oxygen ordering in GdO$_{0.5}$ planes, sketched in Fig. 3(a). The
alternating filled and empty oxygen chains create two types of structural
environment, octahedral and pyramidal, for Co ions; one of these
environments stabilizes the LS ground state ($t^6_{2g}$, $e^0_g$; $S=0$),
while the other one makes the IS state ($t^5_{2g}$, $e^1_g$; $S=1$)
preferable. While our experiments can hardly distinguish whether the IS
ground state is realized in pyramidal or octahedral positions, a
structural study by Frontera {\it et al.} \cite{Frontera} suggests the IS
state for Co ions in pyramidal sites. In this case, magnetic Co ions form
2-leg ladders extended along the $a$ axis [black spheres in Fig. 3(a)] and
separated by non-magnetic layers.

\begin{figure}[t]
\includegraphics*[width=20.5pc]{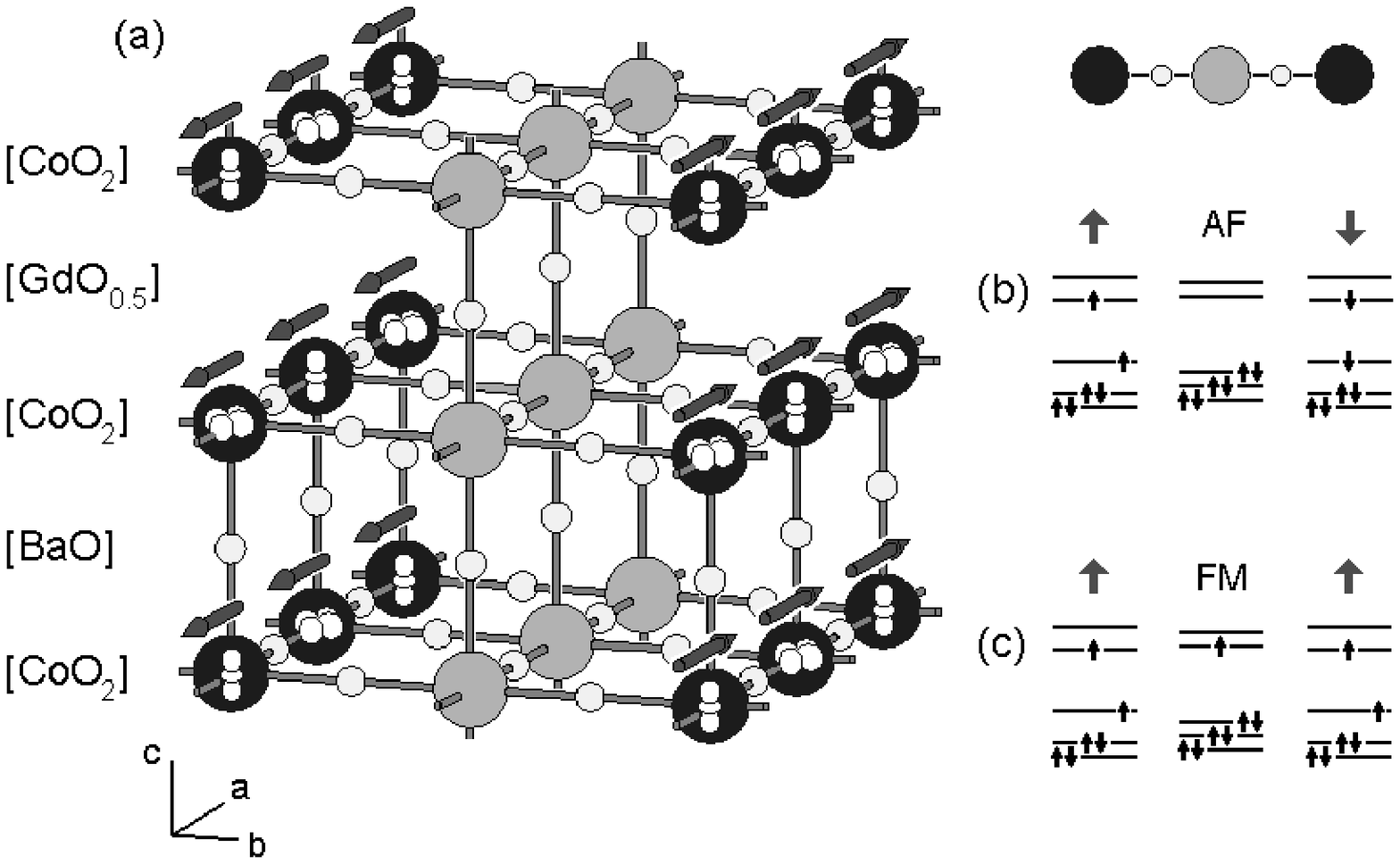}
\caption{(a) A sketch of the crystal and magnetic structure
of GdBaCo$_{2}$O$_{5.5}$. Owing to the ordering of oxygen (small shaded
spheres) into alternating empty and filled chains, the Co ions become
nonequivalent and exhibit either IS state (black spheres, where occupied
$d_{3z^2-r^2}$ or $d_{x^2-y^2}$ $e_g$ orbitals are shown), or LS state
(shaded spheres). Ba and Gd are omitted for clarity. (b) The $b$-axis
interaction of IS-Co$^{3+}$ ions mediated by LS-Co$^{3+}$ results in AF
ordering. (c) Excited (Co$^{2+}$) ions switch the $b$-axis interaction
into the FM one.}
\label{fig3}
\end{figure}

The magnetic ordering in insulators is known to be predominantly caused by
the superexchange (SE) interaction, whose sign for each pair of ions can
be estimated using Goodenough-Kanamori rules \cite{Anderson}. The FM spin
order in ladders can be explained by the orbital ordering among IS
Co-ions, an example of which is shown in Fig. 3(a). Owing to the reduced
dimensionality (quasi-1D/2D) of the ladders, the FM order develops quite
gradually upon cooling, being subject to strong fluctuations, as is
evident in the $M(H)$ curves in Fig. 2(a). Eventually, FM-ordered ladders
with spins aligned along the $a$ axis are formed; however, whether a
macroscopic magnetic moment will emerge or not depends on the relative
orientation of moments between these ladders, which can be {\it only}
ferro- or antiferromagnetic due to the Ising nature of the spins.
Experimentally, the interaction between different FM ladders, mediated by
spinless-Co-O layers [Fig. 3(b)], turns out to be antiferromagnetic,
bringing about the AF ground state. Of course, our magnetization data
cannot tell along exactly which axis, $b$ or $c$, or both, the ladder
stacking is AF, but this does not matter for the qualitative picture.

Quite naturally, the inter-ladder coupling is weak, and switching from the
ground-state AF order to the FM one can be induced by magnetic fields or
temperature. To understand the role of temperature, one should consider
the thermally-excited states: upon heating, a certain amount of LS
Co$^{3+}$ ions become Co$^{2+}$, Co$^{4+}$, or change their spin state.
Whatever the case, each excited ion acquires a non-zero spin, and replaces
the weak SE interaction between two IS Co-ions on the neighboring ladders
with usual SE interactions, providing a strong bridge between spin-ordered
ladders (Fig. 3(c)). Regardless of whether the excited spin couples ferro-
or antiferromagnetically with the spins in ladders, the symmetry of the
bridge makes it certain that the additional coupling between the ladders
is FM. Therefore, thermally-excited spins should inevitably induce an
AF-FM transition at some temperature, if fluctuations will not kill the FM
order in the ladders first. It is worth noting also that the AF ordering
can be suppressed not only by magnetic fields or thermal excitations, but
by changing the oxygen stoichiometry as well: any deviation from $x=0.5$
also introduces Co$^{2+}$ or Co$^{4+}$ ions and should shift the AF-FM
phase to lower temperature, which we have indeed observed experimentally
\cite{unpubl}.

\begin{figure}[t]
%\leftskip2pt
\includegraphics*[width=20.5pc]{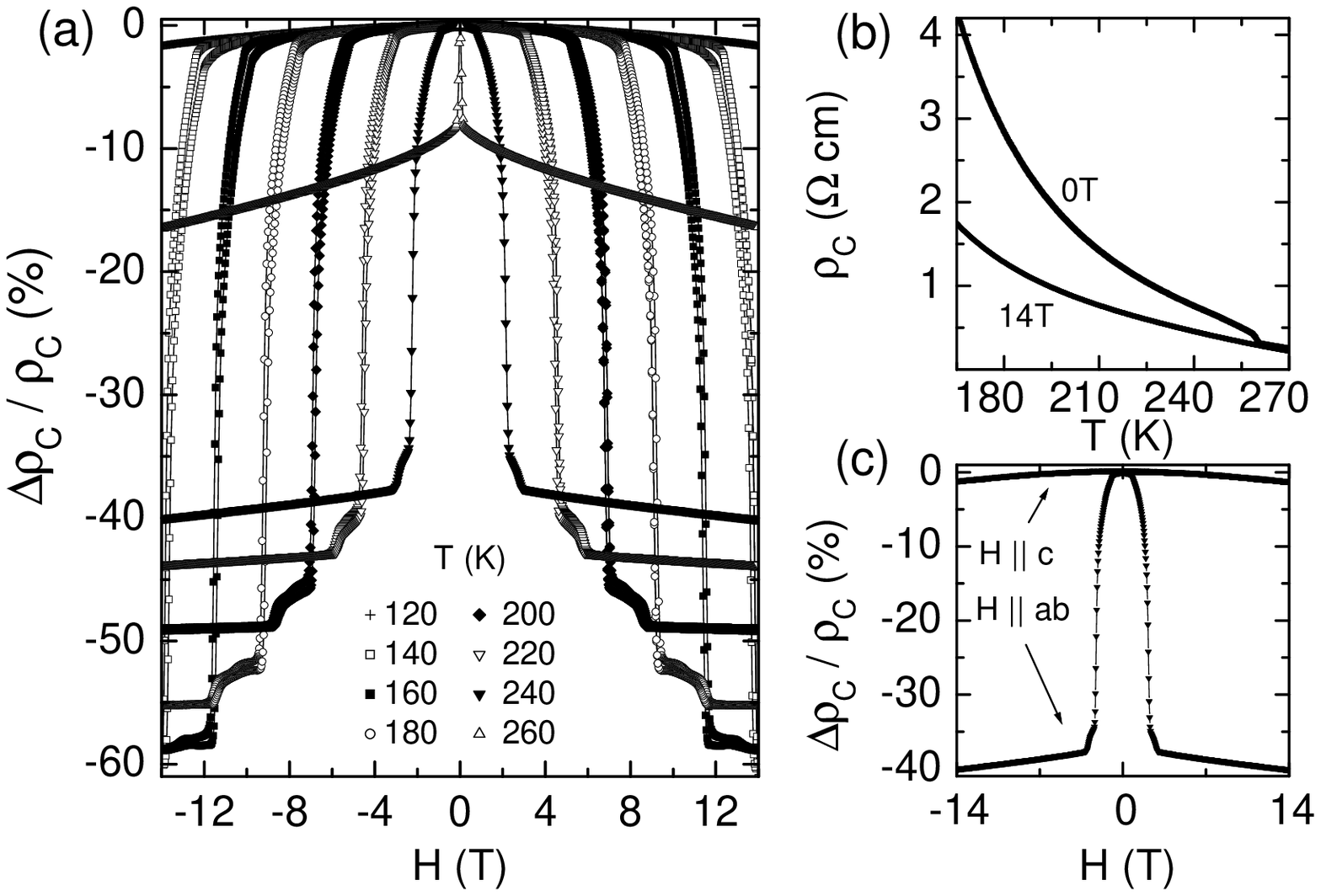}
\caption{$c$-axis MR of a twinned GdBaCo$_{2}$O$_{5.5}$ crystal.
(a) $\Delta\rho_c/\rho_c$ measured at several temperatures for $H\parallel
ab$. (b) $T$-dependence of $\rho_c$ at $H=0$ and 14 T applied along the
$ab$ plane. (c) $\Delta\rho_c/\rho_c$ measured at $T=240$ K for
$H\parallel ab$ and $H\parallel c$.}
\label{fig4}
\end{figure}

To complete the picture of magnetic ordering in GdBaCo$_{2}$O$_{5.5}$, it
is useful to estimate the scale of different magnetic interactions. The
strongest one is definitely the FM superexchange between IS Co$^{3+}$-ions
within ladders, which can be estimated from the Curie temperature $T_c$:
The molecular-field theory for 3D systems would give $J/k_{\text{B}}\sim
150$ K for $T_c=300$ K; however, for the quasi-1D/2D magnetic ordering,
$J$ should be roughly twice as large \cite {Fisher}. The weak AF coupling
between adjacent FM ladders, $J_w$, can be directly evaluated from the
critical field of the AF-FM transition: $J_w/k_{\text{B}}$ reaches $\sim
15$ K at $T=0$, where $H_c\sim 20$ T (Fig. 2(c)). In fact, it is this weak
AF coupling that makes possible the competition and easy switching between
the AF and FM orders in GdBaCo$_{2}$O$_{5.5}$. Lastly, based on the
magnetization anisotropy in the FM or AF state, we can estimate the spin
anisotropy energy, which appears to be unusually large: The rotation of a
spin from the spin-easy $a$ axis to the $b$ axis requires $\Delta
E_{a-b}/k_{\text{B}}\sim 15-20$ K at $T=260$ K, which grows up to $\sim
80-100$ K at $T=2$ K; the energy necessary to turn a spin along the $c$
axis seems to be at least two times larger. Thus, the spin anisotropy
energy appears to be about ten times larger than $J_w$, which explains the
observed Ising-like behavior: the magnetic fields capable of inducing the
AF-FM transition are still too weak to rotate the spins away from the $a$
axis.

One might wonder how the reorientation of weakly coupled ladders affects
the charge transport, bringing about a giant magnetoresistance. The MR
mechanism may be quite simple: GdBaCo$_{2}$O$_{5.5}$ appears to be a
narrow-gap insulator, where the carrier generation goes through formation
of Co$^{2+}$ - Co$^{4+}$ pairs, and the excitation energy for these states
may well depend on the magnetic order. Indeed, in a low-spin Co-O layer
($\parallel ac$), each thermally-excited state strongly couples with two
adjacent spin-ordered ladders, providing them with a FM bridge;
correspondingly, if the ladders' moments are AF oriented, the resulting
frustration should increase the energy of the excited state by $\sim 2J$.
Thus, the relative ordering of adjacent FM ladders is capable of
significantly changing the insulating-gap size. Applied magnetic fields
align the FM ladders and reduce the insulating gap, which results in
step-like increase in the number of carriers and decrease in resistivity
[Fig. 4(a)]. Figure 4(b) demonstrates that the activation energy is
actually diminished by the magnetic field, and thus the MR grows roughly
exponentially upon decreasing temperature.

Two salient points should be emphasized in the MR behavior. The first one
is the cooperative nature of the MR, whereby an apparently small energy of
the magnetic field $g\mu_{\text{B}}H/k_{\text{B}}$, $\sim$ several K, is
capable of changing the carriers' activation energy by several hundreds K.
The second point is the extremely large MR anisotropy: the 14-T field
$H\parallel c$ can do nothing comparable to the MR caused by $H\parallel
a$ [Fig. 4(c)], which confirms the remarkable Ising-like spin anisotropy,
which could be overcome only by magnetic fields in the 100-T range.

The revealed Ising-like spin behavior together with the absence of
significant structural disorder turn RBaCo$_{2}$O$_{5.50}$ into a model
system for studying the competing magnetic interactions and accompanying
MR phenomena. Moreover, RBaCo$_{2}$O$_{5.50}$ compounds may represent an
intriguing realization of a new MR scheme -- a kind of ``magnetic
field-effect transistor'' -- where the charge-carrier injection into a 2D
semiconducting channel is controlled by a magnetic state of neighboring
``ligands''.

\begin{acknowledgments}
We thank S. Komiya and K. Segawa for technical assistance and I. Tsukada
for fruitful discussions. A.A.T. gratefully acknowledges support from
JSPS.
\end{acknowledgments}

\end{document}